\documentclass{cs20proc}

\usepackage{txfonts}
\editors{ }
\publisher{Zenodo}
\conference{The 20.5th Cambridge Workshop on Cool Stars, Stellar Systems, and the Sun}
\conferencedate{2021}

\title{KIC 2852961 – a superflaring red monster in the \emph{Kepler} field}
\author{Zs. Kővári,$^{1}$ 
        K. Oláh,$^{1}$
        M.N. Günther,$^{2}$
        K. Vida,$^{1}$
        L. Kriskovics,$^{1}$
        B. Seli,$^{1}$}

\affiliation{$^{1}$ Konkoly Observatory, Research Centre for Astronomy and Earth Sciences, Budapest, Hungary \\
			 $^{2}$ Department of Physics and Kavli Institute for Astrophysics and Space Research, Massachusetts Institute of Technology, Cambridge, MA 02139, USA}

\shorttitle{KIC 2852961 – a superflaring red monster}
\shortauthors{Zs. Kővári et al.}

\abs{Superflares on giant stars have up to 100,000 times more energy than the high energy solar flares. However, it is disputed, whether scaling up a solar-type dynamo could explain such a magnitude difference. We investigate the flaring activity of KIC\,2852961, a late-type spotted giant. We seek for flares in the \emph{Kepler} Q0-Q17 datasets by an automated technique together with visual inspection. Flare occurence rate and flare energies are analyzed and compared to flare statistics of different targets with similar flare activity at different energy levels. We find that the flare energy distribution of KIC\,2852961 does not seem to be consistent with that of superflares on solar-type stars. Also, we believe that in case of KIC\,2852961 spot activity should have an important role in producing such superflares.
}

\begin{document}

\maketitle

\section{Introduction}
The high magnetic energy outbursts by flares originate from magnetic reconnection, which presumes an underlying dynamo action, when rotation interferes with convective  motions. The most powerful superflares releasing 10${^6}$ times more energy than the largest X-class solar flares are from red giants \citep{2015MNRAS.447.2714B}. However, after the main sequence, at the red giant branch, slowed down rotation and increased size are expected to result in weaker magnetic fields and therefore lower level of magnetic activity. Then what kind of mechanism could provide sufficient energy for the most powerful superflares on giant stars? In fact, magnetic activity can indeed be strong along the red giant branch, which recently has been verified by direct imaging of large starspots \citep{2016Natur.533..217R}. \citet{2020NatAs...4..658L} demonstrated that a common dynamo scaling can be achieved for late-type main sequence and giant stars only when both stellar rotation and convection are taken into account. This infers that magnetic dynamo action related flares in solar-type stars and superflares, for instance, in late-type giants can be linked by scaling. 

In this study we analyse the full \emph{Kepler} data of the spotted red giant star KIC\,2852961 (2MASS\,J19261136+3803107, TIC\,137220334) in order to unravel the relationship between spot activity and flare energies.
The astrophysical properties of KIC\,2852961 have recently been revised by \citet{2020A&A...641A..83K} using all the available ground based and space-borne observations and spectroscopic data. The stellar parameters of the red giant are summarized in Table\,\ref{table1}.

\begin{table}[thb]
	\centering
\caption{Revised astrophysical data of KIC\,2852961}
	\label{table1}
	\begin{tabular*}{\linewidth}{l @{\extracolsep{\fill}} l l}
	\noalign{\smallskip}\hline\hline\noalign{\smallskip}
  Parameter               &  Value \\
	\noalign{\smallskip}\hline\noalign{\smallskip}
  Spectral type            & G9-K0 III  \\
  Photometric  period [d]   &  $\approx$$35.5$ \\
  $T_{\rm eff}$ [K] &           $4722^{+77}_{-56}$    \\
  Radius [$R_{\odot}$] &      $13.1\pm 0.9$   \\
  $\log g$ [cgs]  &   $ 2.43\pm0.14$ \\
  Mass [$M_{\odot}$]   & $1.7\pm0.3$   \\
  Distance [pc] &  $813\pm{17}$\\
  $M_{\rm bol}$     [mag]    & $0.032\pm0.090$ \\
  Luminosity [${L_{\odot}}$]         & $76.5^{+6.0}_{-6.3}$  \\
  Metallicity [Fe/H] &  $-0.08^{+0.15}_{-0.1}$ \\
  $v\sin{i}$ [km\,s$^{-1}$]  &   $\approx$17.5 \\
  Inclination [$^{\circ}$]  & $70\pm10$ \\
	\noalign{\smallskip}\hline
	\end{tabular*}
\end{table}

\section{The \emph{Kepler} light curve of KIC\,2852961}

\begin{figure*}[ht]
	\centering
   \includegraphics[width=2\columnwidth]{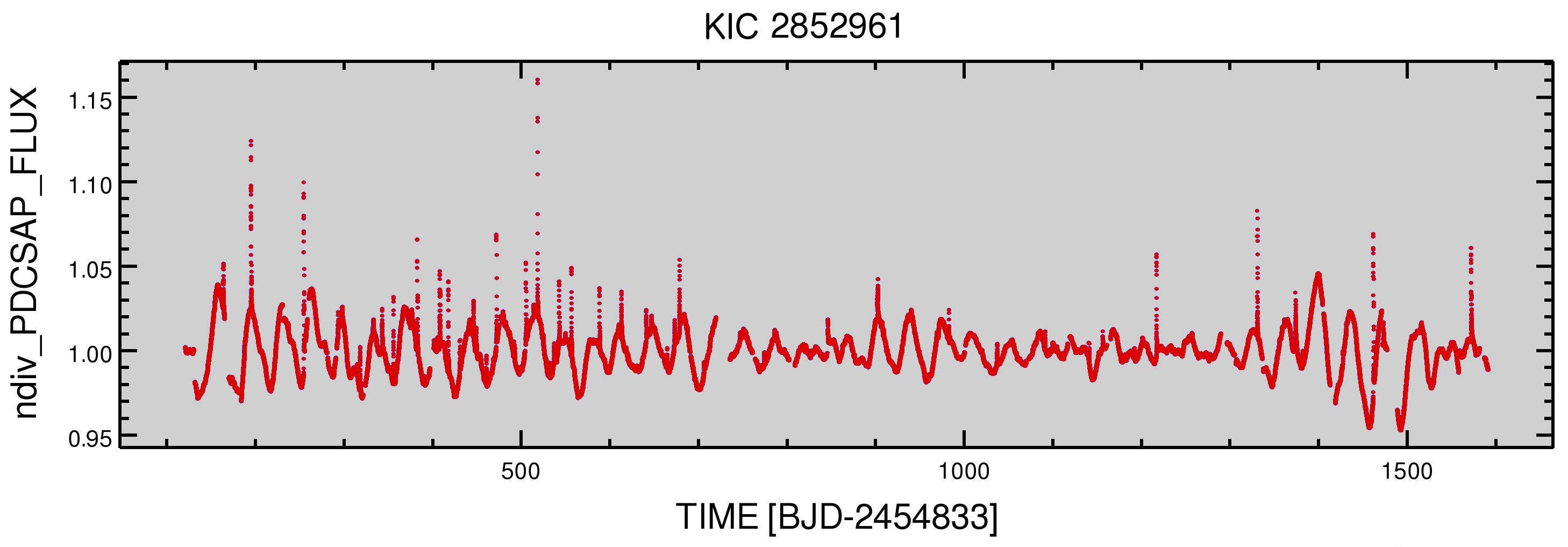}
   \caption{Long-cadence \emph{Kepler} data of KIC\,2852961. The light curve is solidly modulated, typical for a star with constantly renewing spotted surface. Several flare events can also be seen, including very large superflares.}
              \label{fig1}
\end{figure*}

The 18 long cadence Kepler light curves of KIC\,2852961 were collected between BJD\,2454953.5--2456424.0. The normalized data are plotted in Fig.\,\ref{fig1}, indicating the rotational modulation of $\approx$35.5 days due to spots together with strong flare activity.
The light curve has three well defined phases: the first part until BJD=700 (+2454833) with large amplitudes and a lot of energetic flares, the second part from BJD=700 to 1300 with small amplitude and less/less energetic flares, while the third part from BJD=1300 until the end of the dataset with large amplitudes and some energetic flares.

\section{A superflaring monster}

\begin{figure}[bht]
	\centering
   \includegraphics[width=\columnwidth]{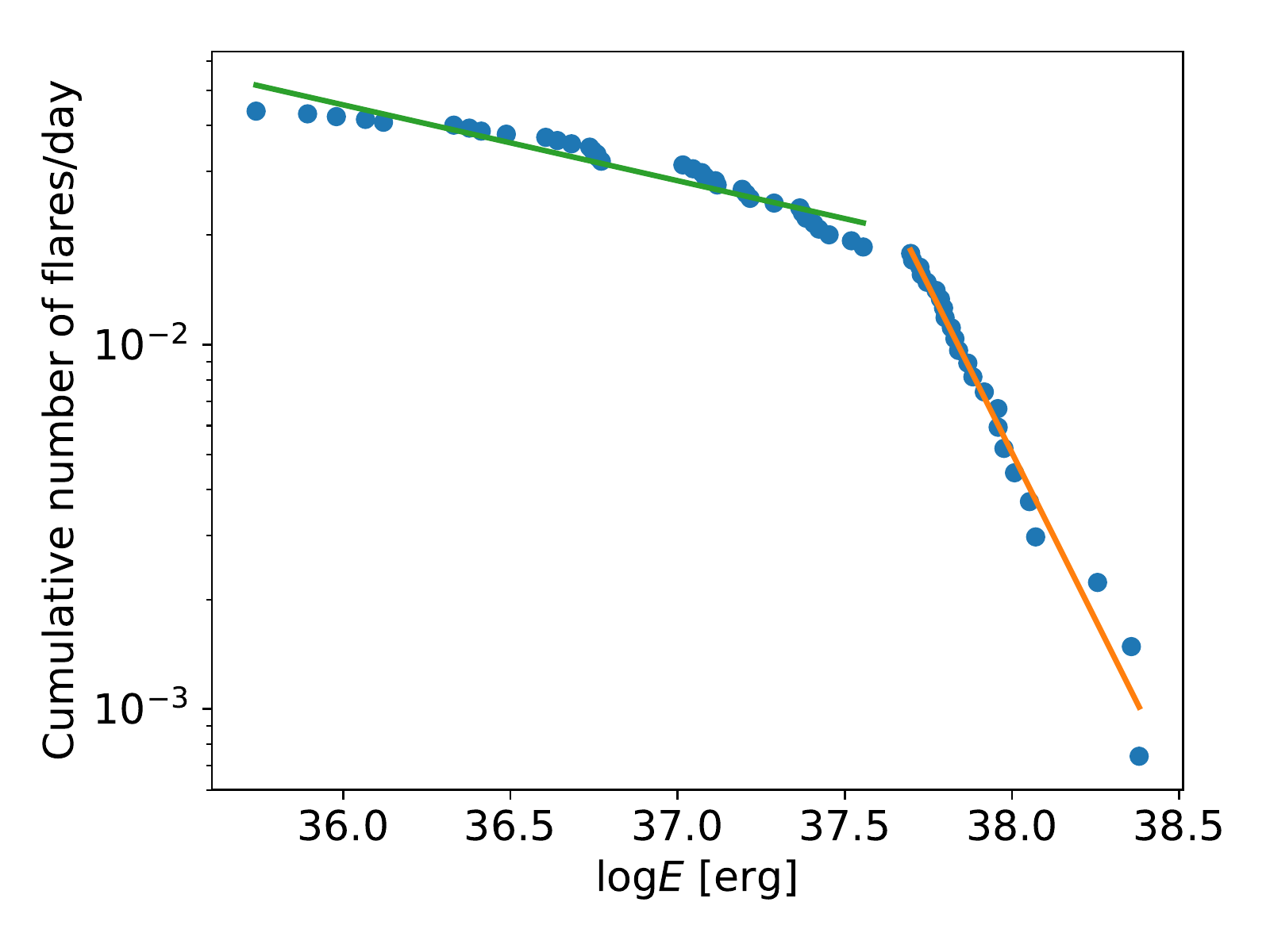}
   \caption{Cumulative flare-frequency  diagram for the red giant KIC\,2852961. The fit to the lower energy range below the breakpoint (green line) yields $\alpha$=1.20 parameter, while the fit for the high energy range above the breakpoint (orange line) gives $\alpha$=2.84.}
              \label{fig2}
\end{figure}

Flares in the \emph{Kepler} light curve of KIC\,2852961 were detected using an automated technique. We applied an updated version of the flare detection pipeline from \citet{2020AJ....159...60G}. To confirm the result we checked the events by visual inspection as well. At the end we confirmed 59 flare events. For each events we calculated flare energies \citep[][]{2020A&A...641A..83K}. The resulting energy values span over three orders of magnitude, ranging between $10^{35}$ and $10^{38}$ ergs. From the flare energies we constructed the cumulative flare frequency distribution, which is shown in Fig.\,\ref{fig2} where flare energies follow
a broken power-law. In this log-log representation the slope of a linear fit is usually defined as $1-\alpha$. The linear fit to the high energy
end  over the range E$\ge$5$\times$10$^{37}$\,erg ("superflare part"; orange line) results in $\alpha$=2.84, which is significantly higher than that
derived for flaring dwarf stars 
\citep[e.g.][etc.]{Howard_2018,2018ApJ...858...55P,2019A&A...622A.133I,2019ApJS..241...29Y}. If the
E$\le$5$\times$10$^{37}$\,erg energy range is fitted (say "normal flares"; see the green line in Fig.\,\ref{fig2}) we get $\alpha$=1.20. This broken distribution has observed in dwarf stars as well, however, with breakpoints at lower energies. According to the model by \citet{2018ApJ...854...14M} the breakpoint is interpreted as "critical energy": when reaching this critical energy, the characteristic flare loop size becomes higher than the local scale height depending on the local field strength and density. This breakpoint, however, is different from star to star, and is not necessarily included in the interval of the actual flare energies.

\begin{figure}[thb]
	\centering
   \includegraphics[width=\columnwidth]{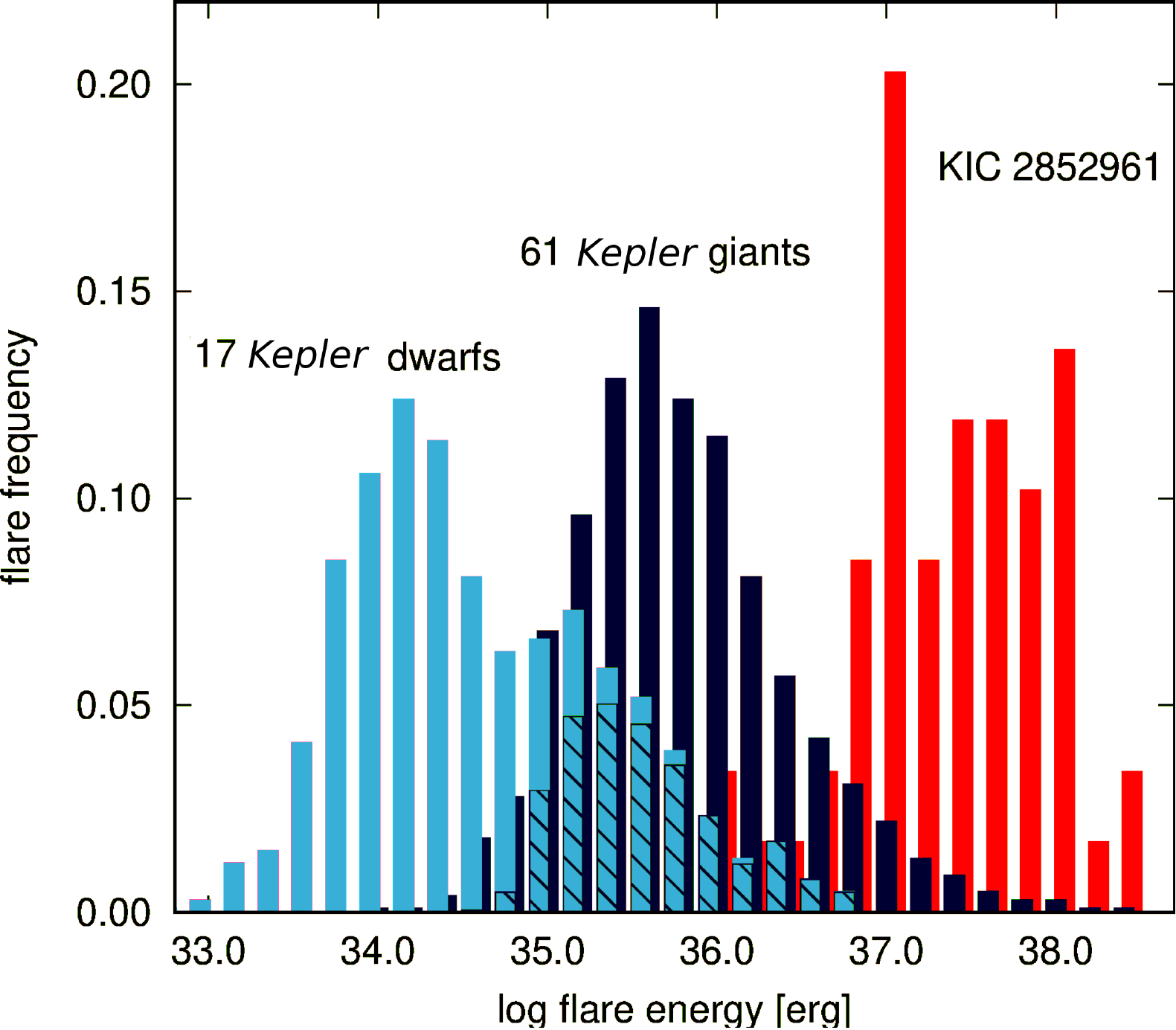}
   \caption{Comparing flare histograms of 17 dwarf stars (light blue) and 61 giant stars (black) from the \emph{Kepler} archive to the flare histogram of KIC\,2852961 (red). Striped part within the dwarf sample corresponds to flares from two subgiants \citep{2020arXiv201007623O}.}
              \label{fig3}
\end{figure}

In Fig.\,\ref{fig3} the flare energy histogram of KIC\,2852961 is compared to those of 17 dwarf (including two subgiants) and 61 giant stars from the \emph{Kepler} field. In this comparison KIC\,2852961 appears as a superflaring red monster with its extremely energetic flares.

\section{The energy source of superflares}

\begin{figure}[thb]
	\centering
   \includegraphics[width=\columnwidth]{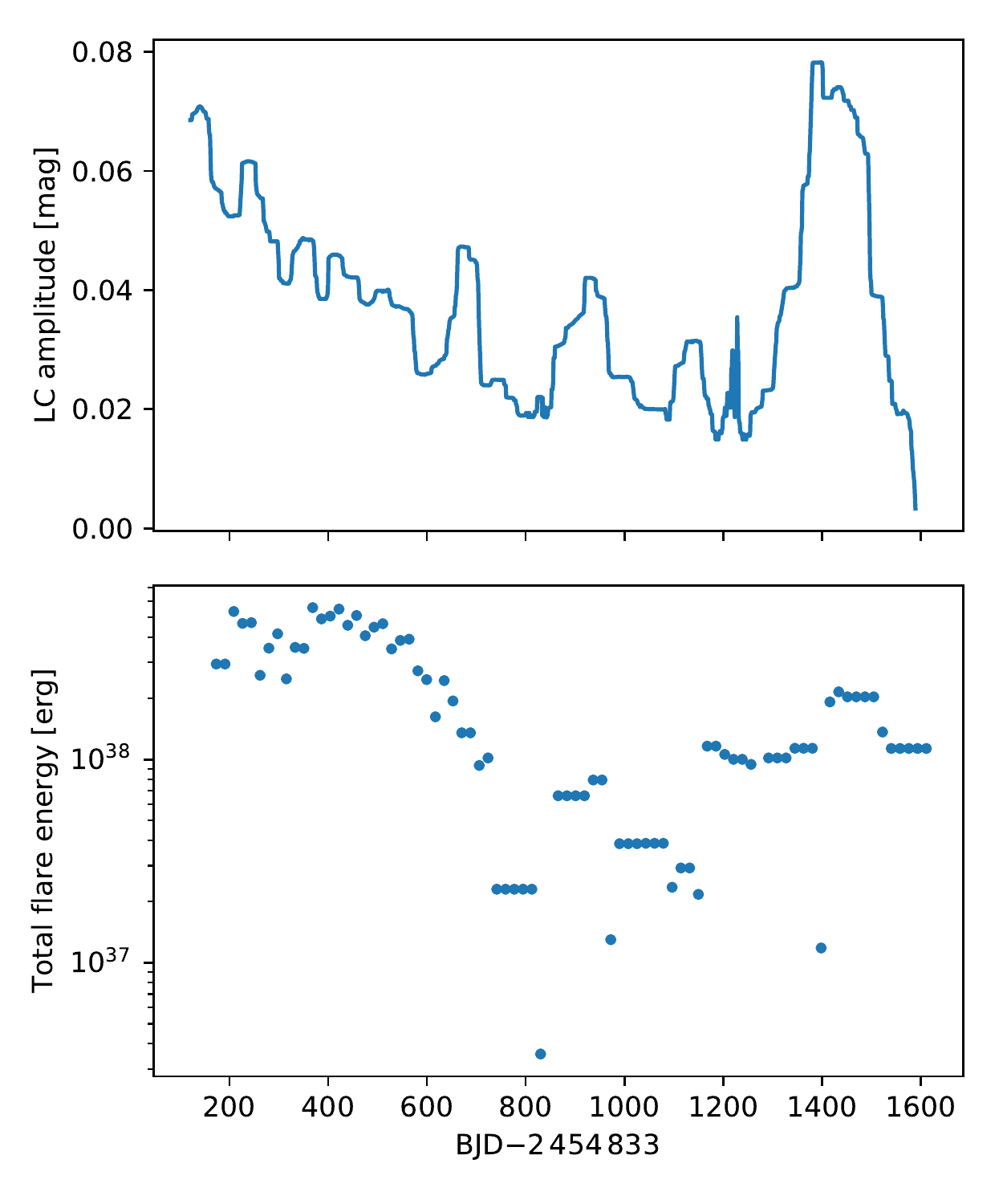}
   \caption{Moving average of the amplitude of rotational modulation in the \emph{Kepler} light curve of KIC\,2852961 (top) and the total flare energy within the same boxcar of $3P_{\rm rot}$ used for the moving average (bottom).}
              \label{fig4}
\end{figure}

As seen in Fig.\,\ref{fig1} during the first term of the whole \emph{Kepler} mission KIC\,2852961 performed large amplitude rotational modulation with a lot of energetic flares. In the middle of the term the amplitude was getting smaller and the flares were getting less powerful and occurred less frequently; at the end of the term the amplitude increased again while flares were more powerful again. In Fig.\,\ref{fig4} we plot the moving average of the overall amplitude change of the light curve cleaned from flares, together with the total flare energy (with applying a boxcar of 3$P_{\rm rot}$). Beyond question, there is a correlation between the rotation amplitude (as an indicator of magnetic activity) and the overall magnetic energy released by flares.
This suggests a general scaling effect behind the production flares in the sense that there are more/more energetic flares when having more/larger active regions on the stellar surface and the flare activity is lower when there are less/smaller active regions.
This is in agreement with the finding of \citet[][see their Figs.\,6-8]{2017PASJ...69...41M}, who demonstrated how the flare-frequency distribution changes with spot sizes on the  Sun and stars: with increasing spot area increasing flare-frequency was found at a certain energy level.

\citet{2015MNRAS.447.2714B} studied \emph{Kepler} flare stars of different luminosities and found that higher luminosity class stars, including giants, have generally higher energy flares. This finding \citep[][see Fig.~10 in that paper]{2015MNRAS.447.2714B} is attributed by the author to a scaling effect, i.e., in the larger active region of a larger star more energy can be stored from the same magnetic field strength, i.e., differences in flare energies of different luminosity class targets are very likely due to size effect.

Taking $T_{\rm spot}=3500$\,K and $\Delta F/F\approx0.04$, maximum relative amplitude from the \emph{Kepler} light curve yields a maximum spot coverage (spot size) of $L\approx0.24 R_{\star}$ \citep[see Eq.~3 in][]{2019ApJ...876...58N}. Assuming $B=3.0$\,kG as a reasonable value of the average magnetic flux density in the spot \citep[cf.][]{2019ApJ...876...58N}, according to the equation in \citet[][see their Eq.~1]{2011LRSP....8....6S}, in the case of KIC\,2852961 the maximum available magnetic energy is estimated to be at least $E_{\rm mag}\approx3.5\times10^{39}$\,erg. However, only a small part of this energy is available to foster flares because it is distributed as potential field energy. Nevertheless, this rough estimation, which is $\approx$15 times higher than the highest flare energies \citep[see][]{2020A&A...641A..83K}, is in fair agreement with the observed flare energies of KIC\,2852961.

\section*{Acknowledgments}
{This work was supported by the Hungarian National Research, Development and Innovation Office grant OTKA K131508, KH-130526 and by the Lendület Program of the Hungarian Academy of Sciences, project No. LP2018-7/2019. Authors from Konkoly Observatory acknowledge the financial support of the Austrian-Hungarian Action Foundation (95\"ou3, 98\"ou5, 101\"ou13). M.N.G. acknowledges support from MIT's Kavli Institute as a Torres postdoctoral fellow.}

\bibliographystyle{cs20proc}
\bibliography{cs205poster_KZS.bib}

\end{document}